\documentclass{aa}
\usepackage{psfig,lscape}
\newcommand{\apj}[2]{ApJ #1, #2}
\newcommand{\aj}[2]{AJ #1, #2}
\newcommand{\aaa}[2]{A\&A #1, #2}
\newcommand{\aas}[2]{A\&AS #1, #2}

\newcommand{\jap}[2]{PASJ #1, #2}
\newcommand{\mon}[2]{MNRAS #1, #2}
\newcommand{\nat}[2]{Nat #1, #2}
\newcommand{\ergs}[1]{$\cdot 10^{#1}$ erg s$^{-1}$}
\newcommand{\hcm}[1]{$\cdot 10^{#1}$ cm$^{-2}$}
\newcommand{\expo}[1]{$\cdot 10^{#1}$}
\newcommand{\lx}{\hbox{L$_{\rm x}$}}
\newcommand{\lmax}{\hbox{L$_{\rm x}^{\rm max}$}}
\newcommand{\et}{et al.}
\newcommand{\ct}{cts s$^{-1}$}
\newcommand{\perr}{r$_{90}$}

\begin{document}
 
\thesaurus{06(11.13.1; %Galaxies: Magellanic Clouds --
              11.19.5; %Galaxies: stellar content --
              08.22.3; %Stars: variables: other --
              13.25.5) %X-rays: stars
          }

\title{Doubling the number of Be/X-ray binaries in the SMC}
 
\author{F. Haberl and M. Sasaki}
\authorrunning{F. Haberl, M. Sasaki}
 
\offprints{F. Haberl (fwh@mpe.mpg.de)}
 
\institute{Max-Planck-Institut f\"ur extraterrestrische Physik,
           Giessenbachstra{\ss}e, 85748 Garching, Germany}
 
\date{Received date; accepted date}
 
\maketitle%\markboth{}
          %         {}

\begin{abstract}
%______________________________________ Do not leave a blank line here!
A correlation of X-ray source and H$_\alpha$ emission-line object
catalogues in the Small Magellanic Cloud (SMC) shows that more
than two thirds of the optically identified Be stars in Be/X-ray
binaries are found as emission-line objects in the catalogues. On
the basis of this result we propose up to 25 X-ray
sources mainly from recent ROSAT catalogues as new Be/X-ray binaries
and give their likely optical counterparts. Also for the five yet 
unidentified X-ray pulsars in the SMC we propose emission-line stars
as counterparts. This more than doubles the number of known
high mass X-ray binary systems in this nearby galaxy. The spatial
distribution of the new candidates is similar to that of the already
identified Be/X-ray binaries with a strong concentration along the SMC
main body and some systems in the eastern wing. The new candidates
contribute mainly to the low-luminosity end of the X-ray luminosity
distribution of Be/X-ray binaries. A comparison with the luminosity 
distribution in the Milky Way reveals no significant differences
at the high-luminosity end and the large number of low-luminosity systems 
in the SMC suggests that many such systems may still be undetected 
in the Galaxy.
The overall ratio of known Be to OB supergiant X-ray binaries in the SMC is
an order of magnitude larger than in the Galaxy, however, might show
spatial variations. While in the eastern wing the ratio is comparable
to that in the Galaxy no supergiant X-ray binary is currently known in
the main body of the SMC. Possible explanations include a different 
star formation history over the last $\sim15$ My.
\end{abstract}
 
%________________________________________________________________
 
\section{Introduction}

In high mass X-ray binaries (HMXBs) a neutron star or black hole 
orbits a massive early-type star and accretes matter either via 
Roche-lobe overflow or from the stellar wind 
which powers the X-ray emission (for recent reviews see Nagase 1989, 
White \et\ 1995, Bildsten \et\ 1997). One divides the class of HMXBs
according to the stellar type of the mass donor star into supergiant
X-ray binaries with luminosity class I-II OB star and Be/X-ray 
binaries with luminosity class III-IV Be star companions. 
Be/X-ray binaries form the larger sub-group of HMXBs. Balmer emission 
lines in the optical spectrum and a characteristic infrared excess are
attributed to the presence of circum-stellar material, probably forming
a disk in the equatorial plane of the Be star.
Be/X-ray binaries often show transient behaviour with two types of 
outbursts. X-ray outbursts repeating with the orbital period are most likely
associated with the passage of the neutron star through the 
circum-stellar disk in an eccentric orbit while giant outbursts, often 
lasting longer than a binary period, probably arise from an expansion
of the disk.

Currently about 100 HMXBs and candidates are known. Nearly one third
were found in the Magellanic Clouds (MCs) from which the majority
is located in the Small Magellanic Cloud (SMC, Coe 1999).
Most of the Be/X-ray binaries in the SMC were discovered in recent years
by X-ray missions like ASCA, BeppoSAX, ROSAT and RXTE (Nagase 1999).
From 20 optically identified HMXBs in the SMC only one is securely 
associated with a supergiant system (the X-ray pulsar SMC\,X--1) 
and from 11 of the 19 Be/X-ray binaries X-ray 
pulsations were detected. Five additional X-ray pulsars are yet to be 
identified, but are most likely also Be systems. The location 
of such a large number of HMXBs at a similar distance makes the SMC
ideally suited for statistical and in particular spatial 
distribution studies of the population of HMXBs in a galaxy as a whole.

Recent surveys to look for H$_\alpha$ emission-line objects 
in the SMC were performed by Meyssonier 
\& Azzopardi (1993, hereafter MA93) and Murphy \& Bessell (1999, MB99).
The survey of MA93 mainly covers the main body and eastern wing of the SMC 
and their catalogue lists 1898 emission-line stars. The catalogue of MB99 
covers nearly all the area where ROSAT PSPC observations of the SMC 
are available 
(except the southern half of the most south-east observation) but is 
less sensitive (372 objects, partially in common with MA93).
A main goal of MA93 and MB99 was to identify planetary nebulae in the 
SMC, however, the catalogues also contain Be stars. MA93 state that all
three at the time of publication known B[e] supergiants which were covered by 
the survey were detected. 
Very few Be/X-ray binaries were known in the SMC until 1993 and
it was not noticed that the Be stars proposed as optical 
counterparts for SMC\,X-3, 2E\,0050.1--7247 and 2E\,0051.1--7304 were 
listed in MA93 as emission-line stars (LIN 198, AzV 111 and AzV 138, 
respectively).
A correlation of the larger sample of Be/X-ray binaries known today in 
the SMC shows that most of them are found in the catalogues of MA93 and 
MB99. In this paper we use the identification of X-ray 
sources with emission-line stars to propose new very likely
candidates for Be/X-ray binaries in the SMC (Sect.~2).
X-ray source catalogues of the SMC which we used for our correlations 
with the emission-line star catalogues were published by Wang \& Wu (1992) 
based on Einstein IPC observations (Seward \& Mitchel 1981, Inoue 
\et\ 1983, Bruhweiler \et\ 1987) and by Haberl \et\ (2000, HFPK00) 
produced from ROSAT PSPC data. We also used a preliminary version of
the ROSAT HRI catalogue of Sasaki \et\ (2000b).

\section{X-ray sources and H$_\alpha$ emission-line stars}

A cross-correlation of the 517 PSPC X-ray sources in the SMC region 
(HFPK00) with the catalogue of H$_\alpha$ emission-line stars published 
by MA93 (1898 entries) yielded 46 possible optical counterparts (distance 
$<$ $ \sqrt{r_{90}^2 + 10^2}$ to account for systematic uncertainties in
X-ray and optical positions, where \perr\ denotes the 90\% statistical
uncertainty of the X-ray position in arc seconds). An additional object
correlating with a PSPC source was found within the catalogue of
candidate emission-line objects of MB99.
From this sample of 47 objects one coincides with a known supernova 
remnant, three with supersoft sources and ten with optically identified 
Be stars proposed 
as optical counterparts for the X-ray sources. Extending the search by
using additional SMC X-ray sources such as from the ROSAT HRI catalogue of
Sasaki \et\ (2000b), the Einstein IPC catalogue of Wang \& Wu (1992) and
X-ray pulsars discovered by instruments on ASCA, BeppoSAX and RXTE yields
another three emission-line stars identified with known Be stars. 
Our correlation results are summarized in Table~\ref{tab-ma93} which is 
sorted in right ascension.
%grouped into PSPC, HRI (not detected by PSPC) and IPC sources (not detected by 
%ROSAT instruments). Within each group the objects are sorted by 
%declination from north to south. 

Columns 2-4 of Table~\ref{tab-ma93} give the source numbers in the X-ray
catalogues for ROSAT PSPC (HFPK00), ROSAT HRI (Sasaki \et\
2000b) and Einstein IPC (Wang \& Wu 1992). For sources detected by ROSAT
coordinates with statistical 90\% error (from HFPK00 when
detected by the PSPC or from Sasaki \et\ 2000b when detected by HRI only)
are given in columns~5-7. For the group of IPC sources which were not 
detected by ROSAT, the IPC coordinates and a 40\arcsec\ error as 
published in WW92 is given. The three digits in column~8 denote the number of
PSPC detections in the energy bands 0.1 - 0.4 keV, 0.5 - 0.9 keV and
0.9 - 2.0 keV. With a few exceptions most of the sources were detected
mainly in the higher energy bands which indicates a hard X-ray spectrum.
HFPK00 used count ratios in the different energy bands, the hardness ratios,
for a spectral classification of the PSPC sources. However, very hard
sources without detection in the lower energy bands were not classified
in HFPK00 because of large errors on the hardness ratios.
Column~9 lists the maximum observed X-ray luminosity for Be/X-ray 
binaries and candidates derived in this work. The values are selected 
from literature or computed from ROSAT count rates using the conversion 
factor 1.67\ergs{37}/\ct\ (see Kahabka \& Pietsch 1996, hereafter KP96), 
typical for X-ray binaries with hard spectrum at the distance of the SMC. 
HRI count rates were converted to PSPC rates using a multiplication 
factor of 3.0 (Sasaki \et\ 2000a). Luminosities derived from count rates 
are indicated with colon.
For all given X-ray luminosities we assume a distance of 65 kpc to the SMC.

Column~10 of Table~\ref{tab-ma93} lists the entry number of the nearest
object in MA93 (MB99 in one case) and column~11 the MA93 classification
type (2 = SNR; 5 = planetary nebula, PN; 9 = late type star). The
distance between X-ray and optical position as listed in MA93 (MB99) is
listed in column~12. B, V and R magnitudes found in the literature 
for identified sources are given in columns 13-15. When available 
B and R obtained from the USNO A2.0 catalogue are listed for the 
remaining sources. In the last column identifications are given 
together with references and new proposals are marked with '?'
behind the source type. 

From the total of 18 high mass X-ray binaries in
the SMC with known Be star as proposed counterpart 13 are found in the
emission-line catalogues of MA93 and MB99. For completeness the
remaining five Be/X-ray binaries are also listed in Table~\ref{tab-ma93}.

The identification of most known Be/X-ray binaries with stars in the
emission-line catalogues of MA93 and MB99 suggests that un-identified X-ray
sources with emission-line star counterparts are most likely also 
Be/X-ray binary systems.
Other objects like SSSs and SNRs can be recognized by their very soft
X-ray spectrum (in contrast to the Be/X-ray binaries with hard spectrum)
or by their X-ray source extent, respectively.
In the following section we summarize the 18 X-ray 
sources optically identified as Be/X-ray binary. We then
propose emission-line stars as likely Be counterparts for the five
un-identified pulsars and in addition for 25 hard X-ray sources. 

\subsection{Supersoft sources}

Three supersoft sources detected by ROSAT were identified with
emission-line objects in the catalogue of MA93. Two of them are
associated with planetary nebulae (PN) while the remaining one is identified
with a symbiotic star in the SMC. More detailed information on the
individual sources and finding charts with X-ray error circles
can be found in Sect.~\ref{sect-notes}.1.

\subsection{Optically identified Be/X-ray binaries}

For 19 X-ray sources in the SMC nearby Be stars were optically identified
and proposed as counterparts, suggesting a Be/X-ray binary nature.
Eighteen were covered by ROSAT observations and information on the 
individual sources is summarized in Sect.~\ref{sect-notes} where also 
finding charts are found. The HEAO source
1H\,0103--762 was not observed with ROSAT (see KP96 and references
therein). Also we do not include the HMXB candidates RX\,J0106.2--7205
(Hughes \& Smith 1994) and EXO\,0114.6--7361 in our summary. For
RX\,J0106.2--7205 no optical spectrum from the suggested counterpart is
published yet, which would confirm its proposed Be star nature. For
EXO\,0114.6--7361 Wang \& Wu (1992) propose the B0\,Ia star AzV\,488 as
counterpart, however, AzV\,477, also a B0\,Ia star is even closer to the
X-ray position. Both candidates suggest a supergiant type of HMXB. It is
remarkable, that together with the only other known supergiant HMXB
SMC\,X--1, EXO\,0114.6--7361 is located in the eastern wing of the SMC.

Fourteen of the identified Be/X-ray binaries were detected by the ROSAT 
PSPC and their X-ray properties can be found in 
HPFK99. AX\,J0051--722, SMC\,X--3 and RX\,J0058.2--7231 
were detected by the ROSAT HRI and are listed in the 
catalogue of HRI sources in the SMC (Sasaki \et\ 2000b).
Only 2E\,0051.1--7304 was not detected by ROSAT.
Thirteen of the proposed Be star counterparts are listed in the catalogues
of MA93 and MB99 and only for five X-ray sources the Be counterparts have
no entry in MA93 and MB99 (AX\,J0049--729, SMC\,X--2, RX\,J0032.9--7348, 
RX\,J0058.2--7231 and XTE\,J0111.2-7317). 
RX\,J0032.9--7348 was not covered by the MA93 survey. 

\subsection{Optically unidentified X-ray pulsars}

Five X-ray pulsars in the SMC were reported for which no optical
identifications are published up to day. Four of them were detected by 
ROSAT PSPC and/or HRI, yielding more accurate positions (HFPK00,
Sasaki \et\ 2000b) and for the fifth case, XTE\,J0054--720, several
ROSAT sources are found within the large RXTE error circle.
In or very close to the ROSAT error circles emission-line 
objects from MA93 are found and we propose these as optical counterparts.
Literature, finding charts and other information on the X-ray binary 
pulsars is presented in Sect.~\ref{sect-notes}. 

Most of the Be stars proposed as optical counterparts for
X-ray sources in the SMC, as summarized in the previous section, are found
as emission-line objects in the catalogue of MA93. This strongly 
supports that the unknown emission-line objects within the error circles of the
unidentified pulsars are also Be star counterparts
of the X-ray pulsars forming Be/X-ray binaries.

\subsection{New Be/X-ray binary candidates}

From the correlation of X-ray source and emission-line object catalogues
34 hard X-ray sources were found with an H$_\alpha$ emission-line object
as possible optical counterpart in the X-ray error circle (see 
Table~\ref{tab-ma93}). The 34 X-ray
sources were investigated in detail to obtain more information which can
help to identify the nature of the object. Finding charts and notes 
to the individual sources are compiled in Sect.~\ref{sect-notes}.

Many sources were observed more than once by ROSAT and we looked for
long-term time variability. In the case of the PSPC we used the 0.9
-- 2.0 keV band because of higher sensitivity for hard sources like
Be/X-ray binaries. To combine detections from the different instruments
we convert HRI to PSPC count rates by multiplying with 3.0 and IPC to
PSPC count rates by multiplying with 1.1 (appropriate for a 5 keV
Bremsstrahlung spectrum with 4.3\hcm{21} absorption column density).
Given the uncertainties in the count rate conversions, variability is only
treated as significant above a factor of 3.
None of the sources was observed with sufficient counting statistics
in order to perform a detailed temporal analysis on shorter time scales 
(within an observation) and to detect X-ray pulsations.

We discuss all un-identified X-ray sources with
emission-line object in or close to the error circle in the following 
and indicate very promising candidates for Be/X-ray binaries with
``Be/X?" in the remark column of Table~\ref{tab-ma93}.

\subsection{Chance coincidences}

To estimate the number of false identifications of X-ray sources
with emission-line objects in MA93 we shifted the X-ray positions of the 
sources in an arbitrary direction and cross-correlated again with the 
MA93 catalogue. To get statistically more reliable results this was 
repeated with different distances between 1 -- 10 arc minutes.
For this purpose we used the PSPC catalogue which 
is most complete. After application of our selection criteria for 
accepting Be/X-ray binary candidates (likelihood of existence for the X-ray 
source $>$ 13, no other identification, distance 
$<$ $ \sqrt{r_{90}^2 + 10^2}$) we find on average
about seven expected chance 
coincidences between PSPC sources and emission-line objects in MA93. 
We emphasize that these are mainly caused by the PSPC sources with 
the largest position uncertainties. The PSPC sources in Table~\ref{tab-ma93} 
with the largest errors on the X-ray position and therefore most likely 
chance coincidences are 99, 248, 295 and 404. Indeed three of them were 
rejected as Be/X-ray binary candidates due to the presence of other 
likely counterparts. Similarly PSPC sources 77 and 253 were disregarded.
Other X-ray sources with large position errors in Table~\ref{tab-ma93}
are the IPC sources which were not securely detected with ROSAT. Also here 
two were not regarded as Be/X-ray binary candidates.
For the 25 new Be/X-ray binary candidates we therefore estimate that 
about two to three may be misidentifications, most likely among those 
with position error \perr\ $>$ 15\arcsec. 

\section{ASCA binary pulsar candidates}

The 1st ASCA Catalogue of X-ray sources in the SMC was compiled by
Yokogawa (1999). The sources were classified according to their hardness
ratios and Be/X-ray binaries were detected as sources with hard X-ray
spectrum in the ASCA energy band. From this classification Yokogawa
(1999) proposed eight new Be/X-ray binary candidates (binary pulsar
candidates, BPc). We correlated our list of Be/X-ray binary candidates
with the ASCA catalogue and find for five of the eight BPc a likely
counterpart within 2\arcmin\ (the maximum ASCA position uncertainty). In
addition one out of the nine (probably because of its low flux)
unclassified ASCA sources also correlates with a PSPC Be/X-ray binary
candidate. In Table~\ref{tab-asca} the Be/X-ray binary candidates with
likely ASCA counterpart are summarized. The first three columns show
ASCA source number, classification and observed X-ray luminosity (0.7 --
10 keV, but corrected for the distance of 65 kpc used throughout this
paper) from Yokogawa (1999). The next two columns contain PSPC source
number and distance between ASCA and PSPC position and the last two
columns give the same for HRI sources. The X-ray luminosities observed
by ASCA are generally a factor of 1.2 -- 2.4 higher than the ROSAT
values (Table~\ref{tab-ma93}) which may only partly be explained by
X-ray variability. Finding a systematically higher luminosity with ASCA 
is probably caused by the different sensitive energy bands and/or 
different intrinsic source spectrum. In particular the PSPC count rate 
to luminosity conversion is very sensitive to the assumed absorption.
In the case of ASCA sources 36 (PSPC 279) and 7
(PSPC 468) the ASCA/ROSAT luminosity ratio is much higher than average
(4.2 and 17, respectively), indicating strong flux variability.

\setcounter{table}{1}
\begin{table}
\caption[]{Be/X-ray binaries with likely ASCA counterparts}
\begin{tabular}{rlcrrrr}
\hline\noalign{\smallskip}
\multicolumn{3}{c}{ASCA}    & \multicolumn{2}{c}{PSPC} & \multicolumn{2}{c}{HRI}\\
No  & Class  & \lx\ [erg s$^{-1}$] & No & d [\arcsec]   & No &  d [\arcsec] \\
\noalign{\smallskip}\hline\noalign{\smallskip}
 2  & BPc    & 3.5\expo{35} & 434  & 30.1              & -- & --   \\
 6  & BPc    & 2.9\expo{35} & 511  & 39.8              & 28 & 27.8 \\
 7  & BPc    & 4.6\expo{35} & 468  & 52.1              & -- & --   \\
27  & BPc    & 8.6\expo{35} & 159  & 15.6              & 95 & 19.0 \\
28  & BPc    & 2.6\expo{35} & 220  & 11.3              & 97 & 20.1 \\
36  & UN     & 9.6\expo{34} & 279  & 56.5              & -- & --   \\
\noalign{\smallskip}
\hline
\end{tabular}
\label{tab-asca}
\end{table}

\section{Optical identifications}

An optical identification campaign of a selected sample of hard ROSAT
PSPC sources from the catalogue of HFPK00 was started independently to
the present work (Keller \et\ in preparation). From three of the X-ray
sources presented here, spectra were taken which in all cases revealed a
Be star nature of the proposed counterpart from the MA93 catalogue. This
confirms RX\,J0057.8--7207 (PSPC 136, Sect.~\ref{sect-notes}.4.8) as
Be/X-ray binary and also the proposed counterpart for the pulsar
AX\,J0105--722 (PSPC 163, Sect.~\ref{sect-notes}.3.2, see also
Filipovi\'c \et\ 2000a) as Be star. RX\,J0051.9--7311 (PSPC 424,
Sect.~\ref{sect-notes}.2.8) was independently identified by Schmidtke
\et\ (1999). These results can be taken as further evidence that the 
emission-line objects we propose for counterparts of X-ray sources
are indeed Be stars.

\section{The Be/X-ray binary population of the SMC}

\begin{figure*}
\centerline{
\psfig{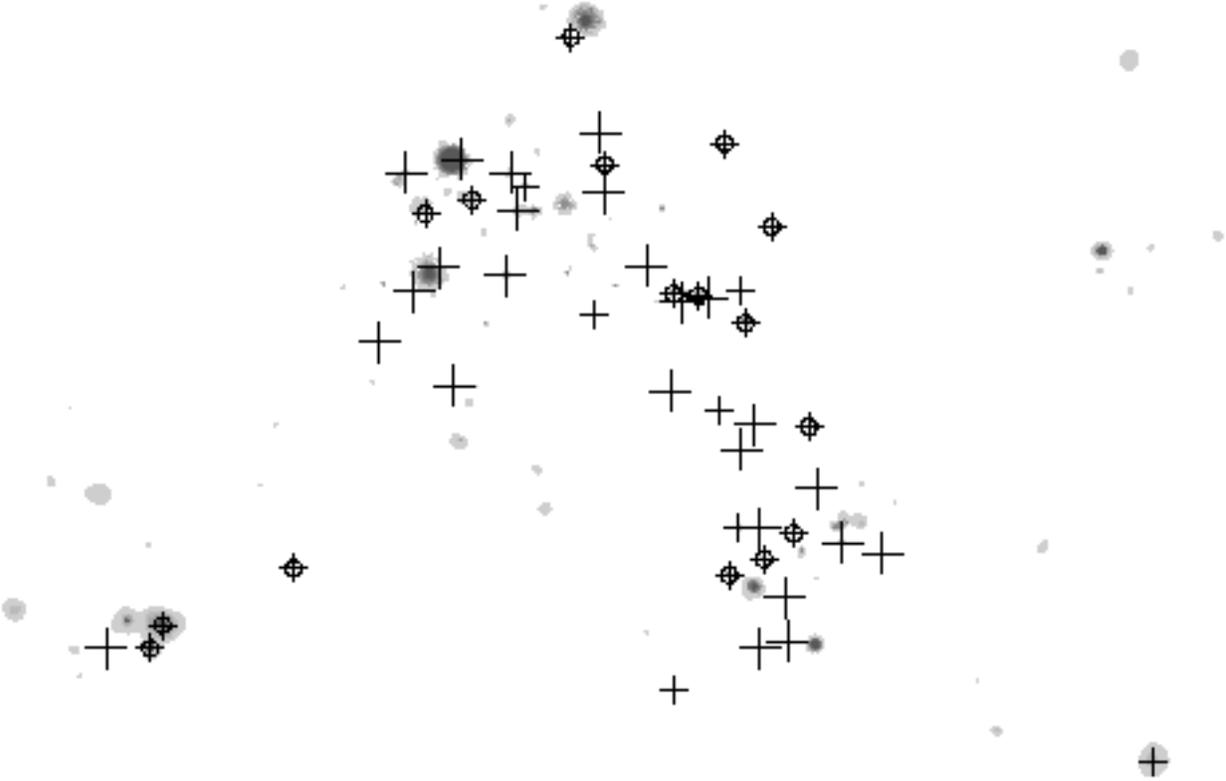}
           }
\caption[]{Distribution of HMXBs in the SMC. X-ray pulsars are marked 
   with circle. The 25 new candidates from this work are indicated by the larger 
   crosses}
\label{fig-distrib}
\end{figure*}

The spatial distribution of the SMC HMXBs including the new candidates 
from this work is shown in Fig.~\ref{fig-distrib}. 
Nearly all new candidates are located along the main body of the SMC
where most of the optically identified Be/X-ray binaries are 
concentrated. Only one new candidate is found in the eastern wing near
the supergiant HMXB SMC\,X--1, where already two other Be/X-ray pulsars are 
known. The distribution is not biased due to incomplete coverage, neither 
in the optical nor in X-rays and makes the strong concentration of 
Be/X-ray binaries in certain areas of the SMC more pronounced. 

\begin{figure}

\centerline{
\psfig{figure=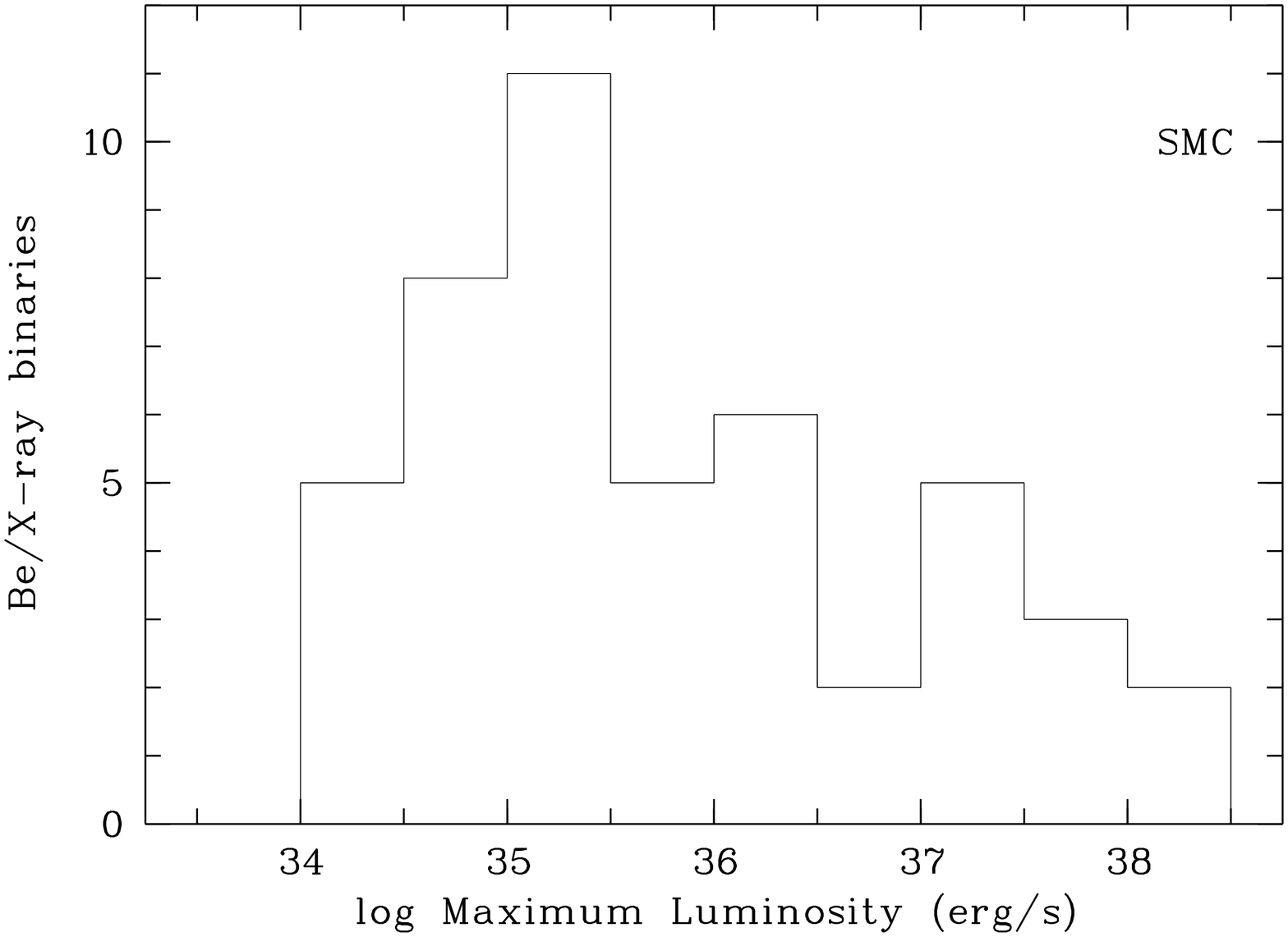,width=88mm,angle=-90,clip=,bbllx=50pt,bburx=540pt,bblly=65pt,bbury=790pt}
           }
\centerline{
\psfig{figure=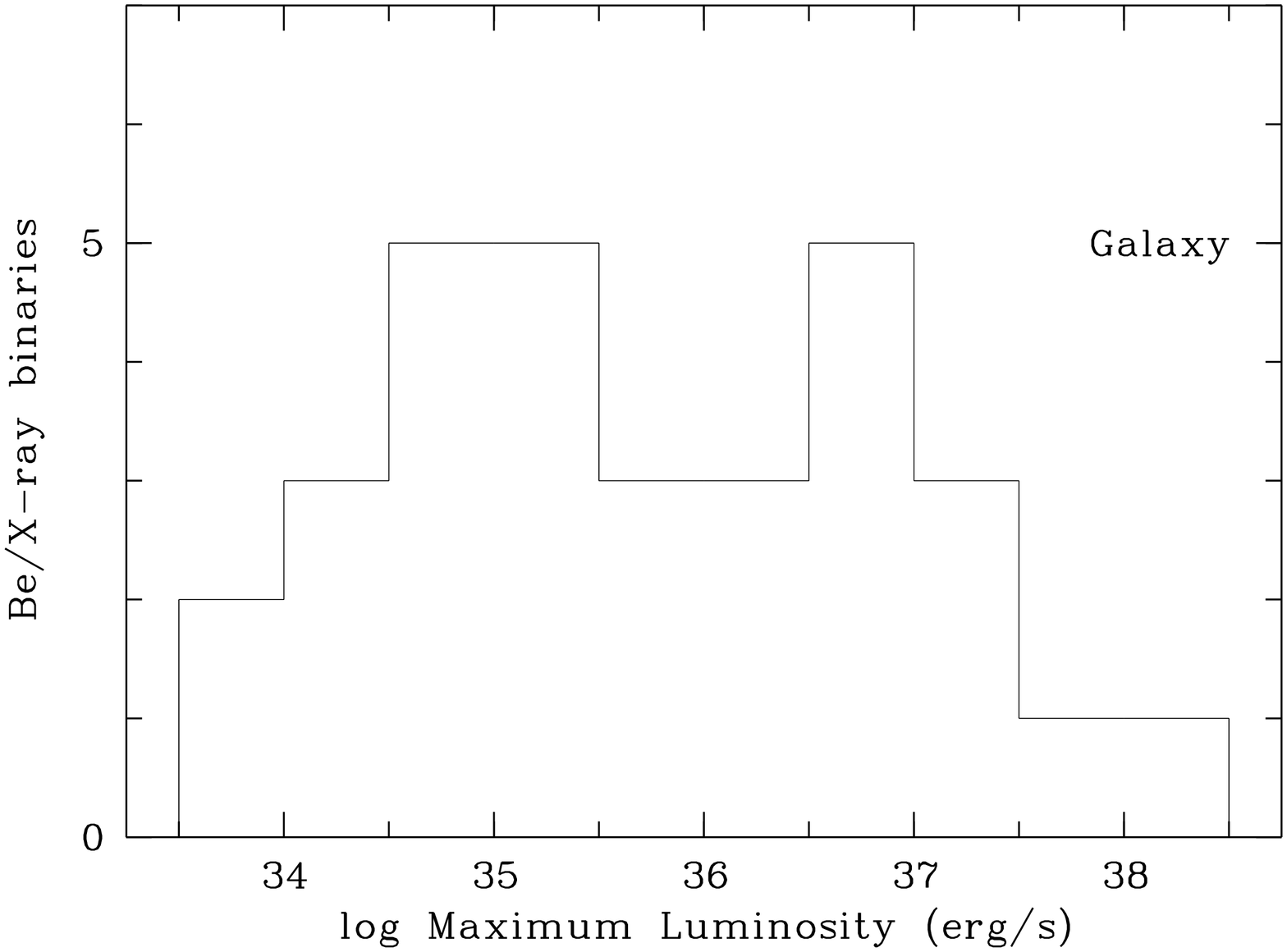,width=88mm,angle=-90,clip=,bbllx=50pt,bburx=570pt,bblly=65pt,bbury=790pt}
           }
\caption[]{Distribution of observed maximum X-ray luminosity of Be/X-ray binaries in the SMC 
   and the Galaxy. Including the candidates from this work, which form 
   a population of low-luminosity systems, 47 Be/X-ray binaries are found 
   in the SMC}
\label{fig-lx}
\end{figure}

The X-ray luminosity distribution of Be/X-ray binaries and candidates in
the SMC is compared to that of systems in the Galaxy in
Fig.~\ref{fig-lx}. To do this we intensively searched the literature on
galactic Be/X-ray binary systems. We derived 31 galactic sources with
\lmax $> 10^{33}$ erg s$^{-1}$ which are summarized in
Table~\ref{tab-begal} (which should be mostly complete). The luminosity
estimates of galactic systems are often hampered by uncertain distances
and different energy bands of the observing instrument. This may cause
luminosity uncertainties by a factor of $\sim$10 in some cases but
should not change the overall distribution drastically.


The new candidates in the SMC mainly raise the number of Be/X-ray binaries
with luminosities log(\lmax) $<$ 35.5 (21 out of 24 are new candidates).
This can easily be explained by the high sensitivity of the ROSAT 
instruments which allowed to detect Be/X-ray binaries in their low-state 
while most of the higher luminosity Be/X-ray binaries were 
discovered during outburst. X-ray luminosities derived from detectors
sensitive at higher energies (typically 0.5 -- 10 keV) might be up to a factor 
$\sim$2 higher than those derived from ROSAT count rates (see Sect.~3) 
which would shift the low-luminosity end in
Fig.~\ref{fig-lx} by 0.3 dex to the right. However, such a shift would 
not change the overall distribution significantly.
Recently, in the Galaxy several likely 
low-luminosity Be/X-ray binaries were discovered by BeppoSAX and ASCA
(1SAX\,J1324.4-6200, Angelini \et\ 1998; 
AX\,J1820.5-1434, Kinugasa \et\ 1998;
AX\,J1749.2-2725, Torii \et\ 1998;
1SAX\,J1452.8-5949, Oosterbroek \et\ 1999; 
AX\,J1700062-4157, Torii \et\ 1999).
ROSAT also contributed new low-luminosity systems (RX\,J0440.9+4431,
RX\,J0812.4-3114, RX\,J1037.5-5647, RX\,J0146.9+6121, Motch \et\ 1997).
However, the high absorption in the galactic plane makes the detection
of low-luminosity X-ray sources in the ROSAT X-ray band and their 
optical identification difficult. This
might explain the smaller number of low-luminosity Be/X-ray binaries
discovered so far in the Galaxy compared to the SMC and might suggest
that the luminosity distribution of Be/X-ray binaries is very similar in
the SMC and our Galaxy. In this case many more such systems are expected to
be found in our Galaxy which would significantly contribute to the hard 
X-ray galactic ridge emission (Warwick \et\ 1985).

Various authors have suggested the existence of a 
population of low-luminosity systems which are usually
persistent X-ray sources showing moderate outbursts and long pulse periods
(e.g. Kinugasa \et\ 1998, Mereghetti \et\ 2000), somewhat different to 
the high-luminosity systems with strong outbursts and shorter pulse periods.
The SMC results suggest that the low-luminosity sources even dominate the
Be/X-ray binaries in number. From the fact that about one third of the 
already identified Be/X-ray binaries is not listed in current emission-line 
catalogues even more such systems are expected to be found in the ROSAT X-ray 
source catalogues of the SMC.
On the other hand some of them will be observed with higher maximum 
luminosity in future outbursts, but if they indeed form a class of 
low-luminosity Be/X-ray binaries like X\,Per, the outbursts are expected 
to be small changing the luminosity distribution immaterial.

There is a large difference in the number of OB supergiant HMXBs
between the Galaxy and the SMC. In the SMC at most two such systems are
identified (SMC\,X-1 and maybe EXO\,0114.6--7361) resulting in an
overall ratio of Be to supergiant X-ray binaries of more than 20. In
the Galaxy this proportion is more of order 2 (12 supergiant systems in 
the Galaxy are listed in the reviews of White \et\ 1995 and Bildsten \et\ 
1997). It is remarkable that the
SMC supergiant HMXBs are all located in the eastern wing giving rise to
a local Be/supergiant HMXB ratio similar to that in the Galaxy. In
contrast no supergiant HMXB is known in the SMC main body making the 
difference more extreme. One possible explanation is a different star
formation history. Be/X-ray binaries evolve from binary
star systems with typical total mass of $\sim$20 M$_{\sun}$
within about 15 My (van den Heuvel 1983) while the more massive
supergiant HMXBs are formed on shorter time scales. The latter therefore
would trace more recent epochs of star formation than the Be/X-ray binaries.
The comparatively large number of Be/X-ray binaries in the SMC 
in this view suggests a
burst of star formation about 15 My ago while relatively few massive
early-type stars were born during the last few million years.

It is remarkable that the Large Magellanic Cloud (LMC) may also have
experienced a burst of star formation about 16 My ago as was derived
from optical photometric surveys by Harris \et\ (1999). LMC and SMC
resemble in the relative composition of their X-ray binary 
populations, both rich in HMXBs but very
few old low-mass X-ray binaries (Cowley \et\ 1999), suggesting a common
star formation history triggered by tidal interaction during close
encounters of LMC, SMC and Milky Way. However, according to present day
modeling the last encounter occurred $\sim$0.2 Gy ago (Gardiner \&
Noguchi 1996), too early for the formation of the Be/X-ray binaries we
see today in X-rays. Therefore, the event which caused the origin of the
frequent SMC (and LMC?) Be/X-ray binaries remains still unclear.
Also the different numbers of HMXBs detected in LMC and SMC relative to 
their total mass need to be explained.

\section{Notes to individual sources}
\label{sect-notes}

In the following notes X-ray source numbers refer to the catalogues of Wang \&
Wu (1992) for Einstein IPC, of HFPK00 for ROSAT PSPC and of Sasaki \et\
(2000b) for ROSAT HRI. Finding charts with X-ray error circles are shown in 
Figs.~\ref{fig-fcsss}, \ref{fig-fcobex}, \ref{fig-fcpbex}, and 
\ref{fig-fcnbex}, for SSSs, already identified Be/X-ray binaries, 
unidentified X-ray pulsars and new Be/X-ray binary candidates, 
respectively. The order within each group of sources follows 
Table~\ref{tab-ma93} and for faster access to the table entry the 
running number is given together with source name.

\subsection{Supersoft sources}

7) RX\,J0048.4--7332:
The SSS RX\,J0048.4--7332 was discovered by Kahabka \et\ (1994) and 
identified as the symbiotic M0 star SMC\,3 by Morgan (1992).
This star is listed in MA93 as object 218 and classified as late type 
star, consistent with the spectral type determined by Morgan (1992).
The accurate HRI position (source 23) confirms the identification
of the PSPC source (512).

39) 1E\,0056.8--7154:
This SSS was discovered in Einstein data (Inoue \et\ 1983) and was
detected with ROSAT PSPC (source 47) and HRI (79). It coincides in
position with the SMC planetary nebula N67 (Aller \et\ 1987) which is listed
as object 1083 in MA93.

41) RX\,J0059.6--7138:
This very soft source was discovered by HFPK00 (PSPC 51) and proposed
as new SSS due to its positional coincidence with the planetary nebula
LIN\,357 (1159 in MA93) in the SMC.

\subsection{Optically identified Be/X-ray binaries}

1) RX\,J0032.9--7348:
Stevens \et\ (1999) identified two Be stars within the PSPC error circle
of RX\,J0032.9--7348, discovered by KP96 as variable source with hard
X-ray spectrum. The a factor of $\sim$5 smaller error radius obtained
from a different PSPC observation by HFPK00 (source 567), however, still
contains both Be stars which are very close to each other
(Fig.~\ref{fig-fcobex}). The two stars were not covered by the survey of
MA93.

9) AX\,J0049--729:
Yokogawa \& Koyama (1998a) reported X-ray pulsations in ASCA data of
this source. Kahabka \& Pietsch (1998) suggested the highly variable
source RX\,J0049.1--7250 (KP96) as counterpart. Stevens
\et\ (1999) identified two Be stars, one only 3\arcsec\ from the X-ray
position and one just outside the error circle given by KP96. 
The revised position of PSPC source 351 in HFPK00 makes the more
distant Be star further unlikely as counterpart (see Fig.~\ref{fig-fcobex}).
None of the two Be stars turns up in the list of emission-line objects of MA93
with the nearest entry (279) 58\arcsec\ away.

13) AX\,J0051--733:
Yokogawa \& Koyama (1998b) discovered X-ray pulsations from this source in
ASCA data. The X-ray source was detected in Einstein IPC, ROSAT PSPC and HRI 
archival data and the 18 year history shows flux variations by at least 
a factor of 10 (Imanishi \et\ 1999).
Cowley \et\ identified already 1997 a Be star as optical
counterpart of the ROSAT HRI source RX\,J0050.8--7316 (HRI 34) which is located
within the ASCA error circle. Cook (1998) reported a 0.708 d period from
this star using data from the MACHO collaboration. The source was also
detected by the PSPC (source 444) and coincides with object 387 in MA93.

16) AX\,J0051--722:
Corbet \et\ (1998b) reported 91 s X-ray pulsations from ASCA observations
of this pulsar which was originally confused with the nearby 46 s pulsar 
XTE\,J0053-724 in XTE data. AX\,J0051--722 was not detected by the PSPC.
An HRI detection reduced the position uncertainty and 
Stevens \et\ (1999) identified a Be star as likely optical counterpart.
The X-ray source is found as source 37 in the HRI catalogue and 
the star is identical to the only emission-line object from MA93 (413) 
in the ASCA error circle.

\begin{table*}
\caption[]{Be/X-ray binaries and likely candidates located in the Galaxy}
\begin{tabular}{lrrll}
\hline\noalign{\smallskip}
Name           & Pulse~~~     & d~~   & \lmax          & Reference for \lmax\\
               & period [s]   & [kpc] & [erg s$^{-1}$] &           \\
\noalign{\smallskip}\hline\noalign{\smallskip}
4U0115+63        &  3.6       &  4    & 3.0\expo{37}  & Tamura \et\ 1992\\
V0332+53         &  4.4       &  3    & 3.2\expo{37}  & Takeshima \et\ 1994\\
2S1553-542       &  9.3       & 10    & 7.0\expo{36}  & Apparao 1994\\
GS0834-430       & 12.3       &  5    & 1.1\expo{37}  & Wilson \et\ 1997\\
XTEJ1946+274     & 15.8       &  5    & 5.4\expo{36}  & Campana \et\ 1999\\
2S1417-624       & 17.6       & 10    & 8.0\expo{36}  & Apparao 1994\\
RXJ0812.4-3114   & 31.9       &  9    & 1.1\expo{36}  & Reig \& Roche 1999a\\
EXO2030+375      & 41.8       &  5    & 1.0\expo{38}  & Parmar \et\ 1989\\
GS2138+56        & 66.3       &3.8    & 9.1\expo{35}  & Schulz \et\ 1995\\
GROJ1008-57      & 93.5       &  2    & 2.9\expo{35}  & Macomb \et\ 1994\\
GS1843-02        & 94.8       & 10    & 6.0\expo{36}  & Finger \et\ 1999\\
4U0728-25        &  103       &  6    & 2.8\expo{35}  & Corbet \et\ 1997\\
A0535+26         &  105       &2.4    & 2.0\expo{37}  & Apparao 1994\\
AXJ1820.5-1434   &  152       &4.7    & 9.0\expo{34}  & Kinugasa \et\ 1998\\
1SAXJ1324.4-6200 &  171       & 10    & 9.8\expo{34}  & Angelini \et\ 1998\\
GROJ2058+42      &  198       &  7    & 2.0\expo{36}  & Wilson \et\ 1998\\
RXJ0440.9+4431   &  203       &3.2    & 3.0\expo{34}  & Reig \& Roche 1999b\\
AXJ1749.2-2725   &  220       &8.5    & 2.6\expo{35}  & Torii \et\ 1998\\
GX304-1          &  272       &2.4    & 1.0\expo{36}  & Apparao 1994\\
4U1145-619       &  292       &0.5    & 7.4\expo{34}  & Apparao 1994\\
4U2206+54        &  392       &2.5    & 2.5\expo{35}  & Saraswat \& Apparao 1992\\
A1118-616        &  405       &  5    & 5.0\expo{36}  & Apparao 1994\\
1SAXJ1452.8-5949 &  437       &  9    & 8.7\expo{33}  & Oosterbroek \et\ 1999\\
AXJ1700062-4157  &  715       & 10    & 7.2\expo{34}  & Torii \et\ 1999\\
X Persei         &  836       &0.83   & 1.9\expo{35}  & Haberl 1994\\
RXJ1037.5-5647   &  860       &5.0    & 4.5\expo{35}  & Reig \& Roche 1999b\\
RXJ0146.9+6121   & 1412       &2.5    & 3.5\expo{35}  & Haberl \et\ 1998\\
A0114+65         &            &2.6    & 1.7\expo{34}  & Motch \et\ 1997\\
Gamma Cas        &            &0.188  & 3.9\expo{34}  & Apparao 1994\\
1E0236.6+6100    &            &3.1    & 2.0\expo{34}  & Motch \et\ 1997\\
1H0521+373       &            &       & 4.0\expo{33}  & Apparao 1994\\
\noalign{\smallskip}
\hline
\end{tabular}
\label{tab-begal}
\end{table*}

19) RX\,J0051.9--7311:
This X-ray source was detected by Cowley \et\ (1997) during ROSAT HRI 
observations of Einstein IPC source 25 and identified with a Be star 
by Schmidtke \et\ (1999). It is identical to PSPC source 424 and HRI 41.
The Be star is found as object 504 in MA93.

20) RX\,J0051.8--7231:
This source was reported as strongly X-ray variable by KP96 
and is associated with the X-ray pulsar 2E\,0050.1--7247 (Israel \et\ 1997).
Observed X-ray luminosities range between 5\ergs{34} and 1.4\ergs{36} 
(Israel \et\ 1997).
The star AzV\,111 (object 511 in MA93) was proposed as counterpart 
for 2E\,0050.1--7247 while Israel \et\ (1997) identified another
H$_\alpha$ active star within their error circle of RX\,J0051.8--7231
which is larger than that of KP96.
Also the position error given for the corresponding PSPC source 265
by HFPK00 is large. The detection of the source in the PSPC observation
600453p (used by KP96) where the source was bright was
rejected by the semi-automatic analysis of HFKP99 because the detection
was close to the support structure of the detector entrance window. A
careful analysis (and using the latest processed data of 600453p) of
the photons of the source in the detector frame shows, however, that it
moved nearly parallel to the window support structure and that it was not
affected by it. In Table~\ref{tab-ma93} therefore the parameters derived
from this PSPC observation are given. They confirm the results of KP96
with small error circle (see Fig.~\ref{fig-fcobex}). Both AzV\,111 and
star 1 of Israel \et\ (1997) are outside this error circle which, 
however,
contains a Be star identified by Stevens \et\ (1999). This star is found
as object 506 in MA93 and is the most probable counterpart of
RX\,J0051.8--7231.

22) SMC\,X--3:
This long-known X-ray source was not detected by the ROSAT PSPC 
but is included in the HRI
catalogue as source 43. The Be star counterpart
(e.g. Crampton \et\ 1978) corresponds to object 531 in MA93.

23) RX\,J0052.1--7319:
Lamb \et\ (1999) reported X-ray pulsations from the variable source
RX\,J0052.1--7319 (KP96) found in ROSAT HRI and CGRO
BATSE data. Israel \et\ (1999) identified a Be star as likely optical
counterpart. It is found in MA93 as object 552 and was detected as
X-ray source by IPC (29), PSPC (453) and HRI (44). The strong X-ray 
variability by a factor of $\sim$200 between different HRI observations
(Kahabka 2000) strongly supports the identification as Be/X-ray binary.

24) 2E\,0051.1--7304:
For this source, listed as entry 31 in the Einstein IPC source catalogue of
Wang \& Wu (1992), the Be star AzV\,138 (Garmany \& Humphreys 1985) was 
proposed as optical counterpart. AzV\,138 corresponds to object 618 in MA93.
2E\,0051.1--7304 was not detected in ROSAT observations.

25) RX\,J0052.9--7158:
This source was detected as X-ray transient by Cowley \et\ (1997) during
ROSAT HRI observations of Einstein IPC source 32 (the largest circle in
the finding chart of Fig.~\ref{fig-fcobex}. Upper limits derived from
PSPC observations imply flux variations by at least a factor of
$\sim$350 (Cowley \et\ 1997). The strong variability and the hard X-ray
spectrum imply a Be/X-ray transient consistent with the suggested Be
star counterpart (Schmidtke \et\ 1999). The Be star is identical to
object 623 in MA93. The X-ray source was detected by ROSAT (PSPC 94 and
HRI 46, the HRI position is most accurate as indicated by the smallest
error circle in the finding chart of Fig.~\ref{fig-fcobex}) and is
located near the edge of the error circle of XTE\,J0054-720. Due to the
large position uncertainty of XTE\,J0054-720 it is, however, not clear if
they are identical.

28) SMC\,X--2:
The long known Be/X-ray binary SMC\,X--2 was caught in outburst with
0.4 \ct\ by the ROSAT PSPC (source 547, see KP96 and 
references therein). Another PSPC observation yielded an upper limit 
indicating X-ray variability of more than a factor of 670. Optical spectra of
the Be counterpart were taken by e.g. Crampton \et\ (1978). The Be
star is located on the rim of the PSPC error circle (Fig.~\ref{fig-fcobex})
and is not contained in the MA93 catalogue.

31) XTE\,J0055--724:
X-ray pulsations from this source were discovered by RXTE (Marshall \&
Lochner 1998) and confirmed in a SAX observation (Santangelo \et\ 1998).
Santangelo \et\ (1998) also report pulsations from archival ROSAT data
reducing the positional uncertainty. Stevens \et\ (1999) identified a
Be star as optical counterpart which corresponds to object 810 in MA93
and which is inside the error circle of PSPC source 241 and HRI source
58.

38) RX\,J0058.2--7231:
RX\,J0058.2--7231 was detected as weak HRI source by Schmitdke \et\ 
(1999) and identified with a Be star. It is contained in the HRI catalogue 
(source 76) but not found in the PSPC catalogue of HFPK00. The Be star is 
not detected in the emission-line star surveys of MA93 and MB99.

40) RX\,J0059.2--7138:
This transient X-ray pulsar with peculiar soft component in the X-ray
spectrum was discovered by Hughes (1994) during an
outburst with a 0.2 -- 2.0 keV luminosity of 3.5\ergs{37}. The X-ray
source was identified with a Be star by Southwell \& Charles (1996) as 
star 1 in their finding chart which is identical to the emission-line object
179 in MB99.

42) RX\,J0101.0--7206:
This source was suggested as X-ray transient by KP96
with a flux variability of at least a factor of 30. 
Stevens \et\ (1999) identified a Be star as optical counterpart.
Object 1 in their Fig.\,1f corresponds to entry 1240 in MA93.

49) SAX\,J0103.2--7209:
Israel \et\ (1998) reported X-ray pulsations from this source consistent
in position with the Einstein source 1E\,0101.5--7225. They confirm the
Be star suggested as counterpart for the Einstein source by Hughes \&
Smith (1994) as the only object in the SAX error circle showing strong
H$_\alpha$ activity. OGLE observations presented by Coe \& Orosz (2000)
confirm this. The Be star corresponds to object 1367 in MA93 and was
also detected by PSPC (source 143) and HRI (101).

58) XTE\,J0111.2--7317:
Chakrabarty \et\ (1998a) reported X-ray pulsations found in RXTE data
from this source located about 30\arcmin\ from SMC\,X--1. Wilson \& Finger
(1998) confirmed the pulsations from CGRO BATSE data and Chakrabarty 
\et\ (1998b) derived an improved position from ASCA data. Two Be stars 
were identified
by Israel \et\ (1999) within or near the ASCA error circle of
30\arcsec. The closer of the two was concluded as most likely counterpart
of XTE\,J0111.2--7317 by Coe \et\ (1999). This Be star has no counterpart 
in MA93. A week source with existence likelihood of 14.5 is 
found in the PSPC catalogue (446). The large error circle of 61\arcsec\ 
overlaps with the ASCA one and includes the position of the Be star. 
There is an additional MA93 object (1731)
within the RXTE error circle and the second Be star found by Israel \et\ 
(1999) is identical to object 1747 in MA93 but both are outside the ASCA
and PSPC confidence regions (see Fig.~\ref{fig-fcobex}).

59) RX\,J0117.6--7330:
Similar to the previous source X-ray pulsations were discovered from the
X-ray transient RX\,J0117.6--7330 (Clark \et\ 1997) in ROSAT PSPC and CGRO
BATSE data (Macomb \et\ 1999). Between two PSPC observations, about 8 
months apart, the count rate diminished by a factor of 270.
Clark \et\ (1997) identified a Be star
counterpart which is identical to object 1845 in MA93 and also within
the error circle of X-ray source 506 in the SMC PSPC catalogue.

\subsection{Optically unidentified X-ray pulsars}

10) AX\,J0049--732:
AX\,J0049--732 was discovered as X-ray pulsar by Imanishi \et\ (1998).
Filipovi\'c \et\ (2000b) reported two hard X-ray point sources from the
catalogue of HFPK00 as possible counterparts to the ASCA pulsar. They suggest
one of them (PSPC source 427) as the more likely counterpart due to its
identification with an emission-line object in MA93 (number 300).

25) XTE\,J0054--720:
The position of this X-ray pulsar could only be determined to an accuracy
of 10\arcmin\ radius (Lochner \et\ 1998). There are at least five X-ray
sources detected by the HRI within that circle (labeled 1 through 5 in
Fig.~\ref{fig-fcpbex} which correspond to the catalogue sources 55, 50,
62, 59 and 46, respectively). Object 2, 4 and 5 were also detected by the
PSPC (104, 157 and 94). The southern of the three (also detected by IPC,
36) is proposed as active galactic nucleus (AGN) 
by HFPK00 and the northern (PSPC 94, HRI 46, IPC
32?) was identified as Be/X-ray transient RX\,J0052.9-7158 (see
Sect.~\ref{sect-notes}.2.2). The Be star counterpart of RX\,J0052.9-7158
coincides with object 623 in MA93. It is not clear if this Be/X-ray 
binary is identical to the RXTE pulsar. A final identification requires 
the detection of pulsations from RX\,J0052.9-7158.

27) XTE\,J0053--724:
Corbet \et\ (1998a) discovered this pulsar and report a ROSAT source 
within the error box. The pulse period, originally confused with 
AX\,J0051--722, was clarified by Corbet \et\ (1998b).
HFPK00 give source 242 as likely counterpart of XTE\,J0053--724. A single
emission-line object from MA93 (717) is found inside the intersecting error
circles of IPC source 34 and the PSPC source.

35) AX\,J0058--720:
X-ray pulsations from this source were discovered by Yokogawa \& Koyama 
(1998b) in ASCA observations. The source was detected in archival 
Einstein IPC, ROSAT PSPC and HRI data which span 18 years and showed 
flux variations by more than a factor of 100 (Tsujimoto \et\ 1999).
This high variability already strongly suggests a Be/X-ray binary.
A single emission-line object from MA93 (1036) is found within the PSPC
error circle (source 114) which is also consistent with the HRI position
(73). It is not clear whether IPC source 41 originates from the same X-ray
source. It may also be associated with another emission-line object 
(1039 of MA93) closer to the IPC position or completely unrelated.

53) AX\,J0105--722:
Yokogawa \& Koyama (1998c) reported AX\,J0105--722 as X-ray pulsar.
From several nearby objects in MA93 number 1517 is closest to the X-ray
position of PSPC source 163. This PSPC source was identified
as likely counterpart of the ASCA pulsar in an area of complex X-ray
emission by Filipovi\'c \et\ (2000a) combining the ROSAT X-ray and radio
data. The star 1517 in MA93 is the northern and bluer component of a pair
of stars close to the error circles of PSPC and HRI detection (110).
The nearby IPC source 53, 77\arcsec\ to the north-east is most likely associated 
to the SNR DEM\,S128 (Filipovi\'c \et\ 2000a).

\subsection{New Be/X-ray binary candidates}

2) RX\,J0041.2--7306:
HFPK00 classified PSPC source 404 as foreground star based on the hardness 
ratios. An emission-line object in the error circle is classified as 
planetary nebula by MA93 indicating a chance positional coincidence.
This makes the identification with the bright star just outside the 
error circle most likely.

3) RX\,J0045.6--7313:
This source (PSPC 436) was detected once in the
0.9 -- 2.0 keV band of the PSPC. An emission-line object in the error
circle suggests an Be/X-ray binary. 

5) RX\,J0047.3--7239:
The PSPC error circle of RX\,J0047.3--7239 (source 295) overlaps with
that of IPC source 19. An emission-line object (168 in MA93 and
classified as late type star) and two radio sources from the catalogue
of Filipovi\'c \et\ (1998) are located in the X-ray confidence region. A
point-like radio source as counterpart would favour an AGN 
identification leaving the nature of RX\,J0047.3--7239 ambiguous.

6) RX\,J0047.3--7312:
RX\,J0047.3--7312 (PSPC 434) is most likely identified with the 
emission-line star 172 in MA93. The fluxes derived from PSPC detections show a
factor of nine variations, supporting that the X-ray source is a Be/X-ray
binary. RX\,J0047.3--7312 is probably identical to IPC source 18, which
showed an intensity within the range observed by the PSPC.
It is also the likely counterpart of ASCA source 2 in Yokogawa (1999; see 
Sect.~3), an X-ray binary candidate detected with similar intensity.

8) RX\,J0048.5--7302:
The emission-line object 238 in MA93 is the brightest optical object in the 
error circle of RX\,J0048.5--7302 (PSPC 392). A Be/X-ray binary is suggested.

11) RX\,J0049.5--7331:
An HRI detection (source 28) with much improved X-ray position compared
to the PSPC (source 511) confirms the identification with the
emission-line object 302 in MA93. RX\,J0049.5--7331 is the probable
counterpart of ASCA source 6 in Yokogawa (1999; see Sect.~3) further
supporting the likely Be/X-ray binary nature.

12) RX\,J0049.7--7323:
This source (PSPC 468) was detected once in the 0.9 -- 2.0 keV band of the 
PSPC. An emission-line object in the error circle suggests an Be/X-ray 
binary. RX\,J0049.7--7323 
is also the likely counterpart of ASCA source 7 in Yokogawa (1999), 
classified as X-ray binary candidate (see Sect.~3).

14) RX\,J0050.7--7332:
RX\,J0050.7--7332 was only once detected by the PSPC (514) and the 
emission-line object in the error circle suggests a Be/X-ray binary 
identification.

15) RX\,J0050.9--7310:
HRI (source 36) and PSPC (source 421) detections are consistent with the
identification of RX\,J0050.9--7310 with the emission-line object 414 in
MA93, suggesting a Be/X-ray binary.

17) RX\,J0051.3--7250:
Two close emission-line objects suggest RX\,J0051.3--7250 (PSPC 349) 
as Be/X-ray binary, but make the identification ambiguous.

18) RX\,J0051.8--7159:
The emission-line object 502 (MA93) found in the error circle of 
RX\,J0051.8--7159 (PSPC 99) is classified as late type star in MA93. 
An active corona of this star may be producing the X-ray emission.
The large error circle contains, however, another bright object which 
could also be 
responsible for the X-rays. The nature of RX\,J0051.8--7159 remains 
therefore unclear.

21) WW 26:
Two emission-line objects from MA93 are found near IPC source 26 
(hardness ratio 0.51, WW92). Object 521 is located inside the error 
circle while 487 can not be completely ruled out as counterpart. No 
ROSAT detection could improve on the position. A 
Be/X-ray binary nature is suggested.

26) RX\,J0053.4--7227:
A precise HRI position (source 48 at the rim of the error circle of PSPC 246) 
with the emission-line star 667 (MA93) as 
brightest object in the error circle makes RX\,J0053.4--7227 a likely 
Be/X-ray binary.

29) RX\,J0054.5--7228:
The uncertainty in the position of RX\,J0054.5--7228 (PSPC 248) 
is relatively large 
and six emission-line objects from MA93 are found as possible 
counterparts to the X-ray source. It is therefore a likely Be/X-ray 
binary but the optical counterpart remains ambiguous.

30) RX\,J0054.9--7245:
Precise ROSAT X-ray positions (PSPC 324 = HRI 57) include an 
emission-line star (809 in MA93) with typical Be star magnitudes as brightest
object in the error circles. A factor of five X-ray flux variability
(the source was bright during a HRI observation) strengthens the
identification as Be/X-ray binary.

32) WW 38 = 2E\,0054.4--7237:
An emission-line object (904 in MA93) is found inside the error circle
of IPC source 38 suggesting a Be/X-ray binary. The source was not 
detected by ROSAT.

33) RX\,J0057.2--7233:
This weak PSPC source (270) was marginally detected once in the hard 0.5
-- 2.0 keV band with a likelihood of 10.4. Unlike all other hard sources
in Table~\ref{tab-ma93} it was not detected in the 0.9 -- 2.0 keV band and
therefore is unlikely a Be/X-ray binary.

34) WW 40 = 2E\,0055.8--7229:
The error circle of IPC source 40 contains two emission-line objects 
from MA93. Object number 1021 is identified as Be star AzV\,111 while
1016, located further north, is of unknown type. ROSAT detected an X-ray 
source inside the IPC error circle (HRI 71 and PSPC 117 with consistent 
positions) which, however, is located between the two emission-line 
objects. The relation between the ROSAT and the Einstein source and the 
emission-line objects is unclear. IPC, HRI and PSPC count rates are 
consistent within a factor of two, which may indicate that they come 
from the same X-ray source. However, the accurate ROSAT positions
make an association with one of the nearby objects from MA93 unlikely.

36) RX\,J0057.8--7207:
Again small error circles from ROSAT HRI (source 74) and PSPC (source
136) observations make the identification of RX\,J0057.8--7207 with an
emission-line star (1038 in MA93) very likely. PSPC detections with
factor of eight different intensities and an HRI detection during an
X-ray bright state which increases the variability to a factor of about
37, make a Be/X-ray binary nature highly probable.

37) RX\,J0057.9--7156:
Be/X-ray binary candidate from positional coincidence of PSPC source 87 
with emission-line object 1044 in MA93 which shows typical optical brightness.

43) RX\,J0101.3--7211:
PSPC detections of this source (PSPC 159 = HRI 95) indicate flux
variations by at least a factor of 15 and the source was not detected in
other observations (upper limit a factor of 100 below the maximum count
rate). This high variability and the presence of an emission-line star
(1257 in MA93) in the small X-ray error circles likely exclude any other
explanation than a Be/X-ray binary. It also is the likely counterpart of
ASCA source 27 in Yokogawa (1999; see Sect.~3), an X-ray binary
candidate.

44) RX\,J0101.6--7204:
Two accurate positions from HRI (source 96) and PSPC (source 121) 
observations suggest the 
identification of RX\,J0101.6--7204 with object 1277 in MA93.
The factor of three variability supports a Be/X-ray binary nature
of RX\,J0101.6--7204 which is probably identical to the IPC source 46 in 
WW92.

45) RX\,J0101.8--7223:
RX\,J0101.8--7223 (PSPC 220 = HRI 97) 
shows X-ray flux variations of a factor of three.
The emission-line star 1288 (MA93) exhibits magnitudes typical for
a Be star in the SMC and is located near the overlapping area of HRI and 
PSPC error circles. We suggest RX\,J0101.8--7223 as Be/X-ray binary as 
it is also the probable counterpart of ASCA source 28 in Yokogawa (1999; see 
Sect.~3), an X-ray binary candidate.

46) RX\,J0102.8--7157:
This weak PSPC source (92) was only once marginally detected in the
broad 0.1 -- 2.4 keV band. The low detection likelihood of 10.5 and the
non-detection in the hard bands indicates that it may not be real, or is
at least not a hard source. A Be/X-ray binary nature is therefore 
unlikely.

47) WW 49:
The IPC source 49 (WW92 give a hardness ratio of 0.21) 
contains a faint emission-line object (1357 in MA93) 
classified as planetary nebula. The spectral hardness of the IPC source 
is inconsistent with an SSS interpretation. The positional coincidence 
is likely by chance.

48) RX\,J0103.1--7151:
This source was detected only once by the PSPC (source 77) and the lowest upper 
limit indicates variability by at least a factor of five, suggesting the 
detection of a single outburst. The emission-line object near the rim of 
the PSPC error circle is, however, the optically weakest (see 
Table~\ref{tab-ma93}), unusual in comparison with identified Be/X-ray 
binaries and candidates derived from this work. We therefore do not 
regard RX\,J0103.1--7151 as prime candidate for a Be/X-ray binary.

50) RX\,J0103.6--7201:
Small error circles from HRI (source 105) and PSPC (source 106)
observations make the identification with object 1393 in MA93 very
likely. RX\,J0103.6--7201 shows variability by a factor of three between
the ROSAT observations, consistent with a Be/X-ray binary.
%{\it It is seen in the Chandra image of the nearby SNR ....}

51) RX\,J0104.1--7243:
Two emission-line objects and a radio source from the catalogue
of Filipovi\'c \et\ (1998) close to RX\,J0104.1--7243
(PSPC 317) make the identification somewhat ambiguous. The most likely
identification with emission-line star 1440 in MA93 suggests 
RX\,J0104.1--7243 as Be/X-ray binary.

52) RX\,J0104.5--7121:
This source was not detected by the PSPC but the accurate HRI position
(source 108) includes only the emission-line object 1470 from MA93 as
bright object in the error circle. RX\,J0104.5--7121 is therefore very
likely a Be/X-ray binary.

54) RX\,J0105.7--7226:
An emission-line star (1544 in MA93) in the PSPC error circle (737) 
suggests RX\,J0105.7--7226 as Be/X-ray binary.

55) RX\,J0105.9--7203:
A single bright object (the emission-line star 1557 in MA93) is found
in the small PSPC error circle (source 120), which makes the identification of 
RX\,J0105.9--7203 as Be/X-ray very likely.

56) RX\,J0107.1--7235:
The probable PSPC detection (279) of IPC source 56 improves the X-ray position 
and allows to identify it with the emission-line star 1619 in MA93.
The source was a factor of 10 brighter during the Einstein observation
and is also the likely counterpart of ASCA source 36 in Yokogawa (1999; see 
Sect.~3) detected with a factor $\sim$4 higher intensity.
A Be/X-ray binary nature is likely.

57) RX\,J0109.0--7229:
The emission-line object 1682 in MA93 is classified as planetary nebula.
X-ray sources associated with planetary nebulae appear as SSSs which is 
not compatible with the hard spectrum of RX\,J0109.0--7229 (PSPC 253). 
The positional coincidence of RX\,J0109.0--7229 is therefore by chance
and the nature of the X-ray source is unclear.

60) RX\,J0119.6--7330:
This source (PSPC 501) was detected once in the
0.9 -- 2.0 keV band of the PSPC. An emission-line object in the error
circle suggests a Be/X-ray binary. 

\section{Summary}

We reviewed the identification of eighteen known Be/X-ray binaries in the 
SMC and found that thirteen of them are listed in emission-line object
catalogues of Meyssonier \& Azzopardi (1993) and Murphy \& Bessell (1999).
From a general correlation of SMC X-ray source and H$_\alpha$ emission-line 
object catalogues we propose optical counterparts for the five optically 
unidentified X-ray pulsars and present 25 new Be/X-ray binary candidates
together with their likely optical counterparts. This more then doubles 
the number of know high mass X-ray binary systems in the SMC.

\acknowledgements
%________________________________________ Do not leave a blank line here!
The ROSAT project is supported by the German Bundesministerium f\"ur
Bildung, Wissenschaft, For\-schung und Technologie (BMBF/DLR) and the
Max-Planck-Gesell\-schaft. The finding charts are 
based on photographic data obtained using The UK Schmidt Telescope.
The UK Schmidt Telescope was operated by the Royal Observatory
Edinburgh, with funding from the UK Science and Engineering Research
Council, until 1988 June, and thereafter by the Anglo-Australian
Observatory.  Original plate material is copyright (c) the Royal
Observatory Edinburgh and the Anglo-Australian Observatory.  
The plates were processed into the present compressed digital form with
their permission.  The Digitized Sky Survey was produced at the Space
Telescope Science Institute under US Government grant NAG W-2166.
%_____________________________________________________________________

\begin{figure*}
%\centerline{\hbox{
%\psfig{figure=dss512.ps,clip=,height=6.9cm,bbllx=82pt,bblly=128pt,bburx=555pt,bbury=552pt}
%\psfig{figure=dss47.ps,clip=,height=6.9cm,bbllx=82pt,bblly=128pt,bburx=555pt,bbury=552pt}
%                 }}
%\centerline{\hbox{
%\psfig{figure=dss51.ps,clip=,height=6.9cm,bbllx=82pt,bblly=128pt,bburx=555pt,bbury=552pt}
%                 }}
  \caption[]{Finding charts produced from the digitized sky survey (DSS)
             for the three X-ray SSSs identified with emission-line 
             objects in the catalogue of MA93 (all objects within 
             60\arcsec\ of the PSPC position are marked with squares). 
             The emission-line stars were localized on the DSS frames 
             using the finding charts published by MA93 which required
             typically a few arcsec shift of the coordinates given in 
             MA93.
             The circles with plus sign at the center mark the X-ray 
             90\% confidence regions (including a 7\arcsec\ systematic 
             error) of ROSAT PSPC and HRI positions. The Einstein IPC error 
             circle of 40\arcsec\ for 1E\,0056.8--7154 is shown
             in the corresponding chart
            }
  \label{fig-fcsss}
\end{figure*}

\begin{figure*}
%\centerline{\hbox{
%\psfig{figure=dss567.ps,clip=,height=6.9cm,bbllx=82pt,bblly=128pt,bburx=555pt,bbury=552pt}
%\psfig{figure=dss351.ps,clip=,height=6.9cm,bbllx=82pt,bblly=128pt,bburx=555pt,bbury=552pt}
%                 }}
%\vspace{10pt}
%\centerline{\hbox{
%\psfig{figure=dss444.ps,clip=,height=6.9cm,bbllx=82pt,bblly=128pt,bburx=555pt,bbury=552pt}
%\psfig{figure=dssaxj0051-722.ps,clip=,height=6.9cm,bbllx=82pt,bblly=128pt,bburx=555pt,bbury=552pt}
%                 }}
%\vspace{10pt}
%\centerline{\hbox{
%\psfig{figure=dss424.ps,clip=,height=6.9cm,bbllx=82pt,bblly=128pt,bburx=555pt,bbury=552pt}
%\psfig{figure=dss265.ps,clip=,height=6.9cm,bbllx=82pt,bblly=128pt,bburx=555pt,bbury=552pt}
%                 }}
  \caption[]{Optically identified Be/X-ray binaries in the SMC (annotations as in 
             Fig.~\ref{fig-fcsss}). IPC error circles always have 40\arcsec\ radius. 
             Two error circles for PSPC 265 are derived from different 
             PSPC observations (see text).
             The small circles in the charts of PSPC 567, 351, 547 and HRI 76 which 
             have no entries in the MA93 catalogue, mark the optically identified Be stars
             proposed as counterparts to the X-ray sources
            }
   \label{fig-fcobex}
\end{figure*}
\addtocounter{figure}{-1}
\begin{figure*}
%\centerline{\hbox{
%\psfig{figure=dsssmcx3.ps,clip=,height=6.9cm,bbllx=82pt,bblly=128pt,bburx=555pt,bbury=552pt}
%\psfig{figure=dss453.ps,clip=,height=6.9cm,bbllx=82pt,bblly=128pt,bburx=555pt,bbury=552pt}
%                 }}
%\vspace{10pt}
%\centerline{\hbox{
%\psfig{figure=dss_ww31.ps,clip=,height=6.9cm,bbllx=82pt,bblly=128pt,bburx=555pt,bbury=552pt}
%\psfig{figure=dss94.ps,clip=,height=6.9cm,bbllx=82pt,bblly=128pt,bburx=555pt,bbury=552pt}
%                 }}
%\vspace{10pt}
%\centerline{\hbox{
%\psfig{figure=dss547.ps,clip=,height=6.9cm,bbllx=82pt,bblly=128pt,bburx=555pt,bbury=552pt}
%\psfig{figure=dss241.ps,clip=,height=6.9cm,bbllx=82pt,bblly=128pt,bburx=555pt,bbury=552pt}
%                 }}
  \caption[]{Continued. AX\,J0051--722, SMC\,X--3 and RX\,J0058.2--7231 were not detected by 
             the ROSAT PSPC. The IPC source 2E\,0051.1--7304 was 
             neither detected by HRI nor PSPC
             }
\end{figure*}
\addtocounter{figure}{-1}
\begin{figure*}
%\centerline{\hbox{
%\psfig{figure=dssrxj0058.2-7231.ps,clip=,height=6.9cm,bbllx=82pt,bblly=128pt,bburx=555pt,bbury=552pt}
%\psfig{figure=dss53.ps,clip=,height=6.9cm,bbllx=82pt,bblly=128pt,bburx=555pt,bbury=552pt}
%                 }}
%\vspace{10pt}
%\centerline{\hbox{
%\psfig{figure=dss132.ps,clip=,height=6.9cm,bbllx=82pt,bblly=128pt,bburx=555pt,bbury=552pt}
%\psfig{figure=dss143.ps,clip=,height=6.9cm,bbllx=82pt,bblly=128pt,bburx=555pt,bbury=552pt}
%                 }}
%\vspace{10pt}
%\centerline{\hbox{
%\psfig{figure=dss446.ps,clip=,height=6.9cm,bbllx=48pt,bblly=70pt,bburx=585pt,bbury=545pt}
%\psfig{figure=dss506.ps,clip=,height=6.9cm,bbllx=82pt,bblly=128pt,bburx=555pt,bbury=552pt}
%                 }}
  \caption[]{Continued. 
%             The small circles indicate optically identified Be stars
             The large circles in the case of PSPC 446 show the RXTE 
             (90\arcsec), PSPC (61\arcsec) and ASCA (30\arcsec) errors
             }
\end{figure*}

\begin{figure*}
%\centerline{\hbox{
%\psfig{figure=dss427.ps,clip=,height=6.9cm,bbllx=82pt,bblly=128pt,bburx=555pt,bbury=552pt}
%\psfig{figure=dssxtej0054-720.ps,clip=,height=6.9cm,bbllx=82pt,bblly=90pt,bburx=544pt,bbury=522pt}
%                 }}
%\vspace{10pt}
%\centerline{\hbox{
%\psfig{figure=dss242.ps,clip=,height=6.9cm,bbllx=82pt,bblly=128pt,bburx=555pt,bbury=552pt}
%\psfig{figure=dss114.ps,clip=,height=6.9cm,bbllx=82pt,bblly=128pt,bburx=555pt,bbury=552pt}
%                 }}
%\vspace{10pt}
%\centerline{\hbox{
%\psfig{figure=dss163.ps,clip=,height=6.9cm,bbllx=82pt,bblly=128pt,bburx=555pt,bbury=552pt}
%                 }}
  \caption[]{Proposed Be/X-ray binaries as optical counterparts of 
             unidentified X-ray pulsars in the SMC. The large error circle of
             XTE\,J0054--720 contains 5 HRI sources, 3 of them were detected by 
             the PSPC (note the larger size of the finding
             chart). The southern is a likely AGN and the northern
             is identified with a Be star (PSPC 94, HRI 46, IPC 32?) which correlates 
             with an MA93 object (see enlarged chart in Fig.~\ref{fig-fcobex} 
             for RX\,J0052.9--7158)
            }
  \label{fig-fcpbex}
\end{figure*}

\begin{figure*}
%\centerline{\hbox{
%\psfig{figure=dss404.ps,clip=,height=6.9cm,bbllx=82pt,bblly=128pt,bburx=555pt,bbury=552pt}
%\psfig{figure=dss436.ps,clip=,height=6.9cm,bbllx=82pt,bblly=128pt,bburx=555pt,bbury=552pt}
%                 }}
%\vspace{10pt}
%\centerline{\hbox{
%\psfig{figure=dss295.ps,clip=,height=6.9cm,bbllx=82pt,bblly=128pt,bburx=555pt,bbury=552pt}
%\psfig{figure=dss434.ps,clip=,height=6.9cm,bbllx=82pt,bblly=128pt,bburx=555pt,bbury=552pt}
%                 }}
%\vspace{10pt}
%\centerline{\hbox{
%\psfig{figure=dss392.ps,clip=,height=6.9cm,bbllx=82pt,bblly=128pt,bburx=555pt,bbury=552pt}
%\psfig{figure=dss511.ps,clip=,height=6.9cm,bbllx=82pt,bblly=128pt,bburx=555pt,bbury=552pt}
%                 }}
  \caption[]{Emission-line objects from MA93 in or near the error 
             circles of unidentified X-ray sources as detected by 
             ROSAT PSPC and HRI and Einstein IPC
            }
  \label{fig-fcnbex}
\end{figure*}
\addtocounter{figure}{-1}
\begin{figure*}
%\centerline{\hbox{
%\psfig{figure=dss468.ps,clip=,height=6.9cm,bbllx=82pt,bblly=128pt,bburx=555pt,bbury=552pt}
%\psfig{figure=dss514.ps,clip=,height=6.9cm,bbllx=82pt,bblly=128pt,bburx=555pt,bbury=552pt}
%                 }}
%\vspace{10pt}
%\centerline{\hbox{
%\psfig{figure=dss421.ps,clip=,height=6.9cm,bbllx=82pt,bblly=128pt,bburx=555pt,bbury=552pt}
%\psfig{figure=dss349.ps,clip=,height=6.9cm,bbllx=82pt,bblly=128pt,bburx=555pt,bbury=552pt}
%                 }}
%\vspace{10pt}
%\centerline{\hbox{
%\psfig{figure=dss99.ps,clip=,height=6.9cm,bbllx=82pt,bblly=128pt,bburx=555pt,bbury=552pt}
%\psfig{figure=dss_ww26.ps,clip=,height=6.9cm,bbllx=82pt,bblly=128pt,bburx=555pt,bbury=552pt}
%                 }}
  \caption[]{Continued}
\end{figure*}
\addtocounter{figure}{-1}
\begin{figure*}
%\centerline{\hbox{
%\psfig{figure=dss246.ps,clip=,height=6.9cm,bbllx=82pt,bblly=128pt,bburx=555pt,bbury=552pt}
%\psfig{figure=dss248.ps,clip=,height=6.9cm,bbllx=82pt,bblly=128pt,bburx=555pt,bbury=552pt}
%                 }}
%\vspace{10pt}
%\centerline{\hbox{
%\psfig{figure=dss324.ps,clip=,height=6.9cm,bbllx=82pt,bblly=128pt,bburx=555pt,bbury=552pt}
%\psfig{figure=dss_ww38.ps,clip=,height=6.9cm,bbllx=82pt,bblly=128pt,bburx=555pt,bbury=552pt}
%                 }}
%\vspace{10pt}
%\centerline{\hbox{
%\psfig{figure=dss270.ps,clip=,height=6.9cm,bbllx=82pt,bblly=128pt,bburx=555pt,bbury=552pt}
%\psfig{figure=dss_ww40.ps,clip=,height=6.9cm,bbllx=82pt,bblly=128pt,bburx=555pt,bbury=552pt}
%                 }}
  \caption[]{Continued}
\end{figure*}
\addtocounter{figure}{-1}
\begin{figure*}
%\centerline{\hbox{
%\psfig{figure=dss136.ps,clip=,height=6.9cm,bbllx=82pt,bblly=128pt,bburx=555pt,bbury=552pt}
%\psfig{figure=dss87.ps,clip=,height=6.9cm,bbllx=82pt,bblly=128pt,bburx=555pt,bbury=552pt}
%                 }}
%\vspace{10pt}
%\centerline{\hbox{
%\psfig{figure=dss159.ps,clip=,height=6.9cm,bbllx=82pt,bblly=128pt,bburx=555pt,bbury=552pt}
%\psfig{figure=dss121.ps,clip=,height=6.9cm,bbllx=82pt,bblly=128pt,bburx=555pt,bbury=552pt}
%                 }}
%\vspace{10pt}
%\centerline{\hbox{
%\psfig{figure=dss220.ps,clip=,height=6.9cm,bbllx=82pt,bblly=128pt,bburx=555pt,bbury=552pt}
%\psfig{figure=dss92.ps,clip=,height=6.9cm,bbllx=82pt,bblly=128pt,bburx=555pt,bbury=552pt}
%                 }}
  \caption[]{Continued}
\end{figure*}
\addtocounter{figure}{-1}
\begin{figure*}
%\centerline{\hbox{
%\psfig{figure=dss_ww49.ps,clip=,height=6.9cm,bbllx=82pt,bblly=128pt,bburx=555pt,bbury=552pt}
%\psfig{figure=dss77.ps,clip=,height=6.9cm,bbllx=82pt,bblly=128pt,bburx=555pt,bbury=552pt}
%                 }}
%\vspace{10pt}
%\centerline{\hbox{
%\psfig{figure=dss106.ps,clip=,height=6.9cm,bbllx=82pt,bblly=128pt,bburx=555pt,bbury=552pt}
%\psfig{figure=dss317.ps,clip=,height=6.9cm,bbllx=82pt,bblly=128pt,bburx=555pt,bbury=552pt}
%                 }}
%\vspace{10pt}
%\centerline{\hbox{
%\psfig{figure=dss_h108.ps,clip=,height=6.9cm,bbllx=82pt,bblly=128pt,bburx=555pt,bbury=552pt}
%\psfig{figure=dss737.ps,clip=,height=6.9cm,bbllx=82pt,bblly=128pt,bburx=555pt,bbury=552pt}
%                 }}
  \caption[]{Continued}
\end{figure*}
\addtocounter{figure}{-1}
\begin{figure*}
%\centerline{\hbox{
%\psfig{figure=dss120.ps,clip=,height=6.9cm,bbllx=82pt,bblly=128pt,bburx=555pt,bbury=552pt}
%\psfig{figure=dss279.ps,clip=,height=6.9cm,bbllx=82pt,bblly=128pt,bburx=555pt,bbury=552pt}
%                 }}
%\vspace{10pt}
%\centerline{\hbox{
%\psfig{figure=dss253.ps,clip=,height=6.9cm,bbllx=82pt,bblly=128pt,bburx=555pt,bbury=552pt}
%\psfig{figure=dss501.ps,clip=,height=6.9cm,bbllx=82pt,bblly=128pt,bburx=555pt,bbury=552pt}
%                 }}
  \caption[]{Continued}
\end{figure*}
% to avoid 2 lines of text written at bottom of figures in 2col version:
%~\\~\\~\\~\\~\\~\\~\\~\\~\\~\\~\\~\\

\clearpage
\onecolumn
\landscape
\setcounter{table}{0}
\begin{table}
\caption[]{X-ray sources correlating with emission-line objects in the catalogues of MA93, MB99 or identified Be-stars}
\scriptsize
\begin{tabular}{rrrrccrrcrcrlllp{75mm}l}
\hline\noalign{\smallskip}
1& 2~  &3~    &4~ &     5      &      6    & 7~   &  8~   &   9          &  10   & 11   &12     &  ~13  &  ~14  &  ~15  & ~~~~16     \\
\hline\noalign{\smallskip} 
 &No  &  No & No  &  RA        &  Dec      &\perr & det   & L$_{\rm x}$  & No    & T    & dist  &   ~B  &   ~V  &   ~R  & Remarks    \\
 &RP  &  RH & EI &&\hspace{-15mm}(J2000.0)&[\arcsec]&    & erg s$^{-1}$ & MA    &      & [\arcsec]&    &       &       &\\
\noalign{\smallskip}\hline\noalign{\smallskip}
 1& 567 &  -- & -- & 00 32 56.1 & -73 48 19 & 12.9 &   012 & 1.3\expo{37} &    -- &   -- &   --  &   --  & --    &   --  & Be/X RXJ0032.9-7348, 2 Be stars (KP96, SCB99)   \\ 
 2& 404 &  -- & -- & 00 41 16.4 & -73 06 41 & 35.6 &   100 &     --       &    22 &    5 &  31.5 & 18.4: & --    & 17.0: & [fg Star] GSC 9141.4223 ?                            \\ 
 3& 436 &  -- & -- & 00 45 37.9 & -73 13 54 & 29.4 &   001 & 1.2\expo{35}:&   114 &   -- &  23.4 & 18.3: & --    & 16.9: & Be/X?                                           \\ 
 4& 413 &  -- & 16 & 00 47 12.2 & -73 08 26 &  6.8 &   012 &     --       &   165 &    2 &   4.5 &   --  & --    &   --  & SNR 0045-73.4 (RLG94) 13cm                           \\ 
 5& 295 &  -- & 19 & 00 47 18.4 & -72 39 42 & 49.8 &   001 &     --       &   168 &    9 &  21.1 &   --  & --    &   --  & AGN? Radio SMC B0045-7255 (FHW98) 13cm               \\ 
 6& 434 &  -- & 18 & 00 47 23.4 & -73 12 23 &  4.0 &   012 & 2.4\expo{35}:&   172 &   -- &   1.6 &   --  & --    &   --  & Be/X? [hard]                                    \\ 
 7& 512 &  23 & -- & 00 48 23.1 & -73 31 43 & 24.1 &   620 &     --       &   218 &    9 &  15.1 &   --  & --    &   --  & SSS RXJ0048.4-7332, symbiotic star M0 (KP96, M92)     \\ 
 8& 392 &  -- & -- & 00 48 33.7 & -73 02 24 &  7.6 &   011 & 1.2\expo{35}:&   238 &   -- &   6.5 & 14.6: & --    & 16.9: & Be/X?                                           \\ 
 9& 351 &  -- & -- & 00 49 02.5 & -72 50 52 & 13.7 &   002 & 6.1\expo{36} &    -- &   -- &   --  & 17.01 & 16.92 &   --  & Be/X AXJ0049-729, 74.67 s pulsar (YK98a, KP98, SCB99, CO00)\\ 
10& 427 &  -- & -- & 00 49 29.6 & -73 10 56 &  5.5 &   002 & 4.1\expo{35} &   300 &   -- &   2.6 & 18.1: & --    & 15.6: & Be/X? AXJ0049-732, 9.132 s pulsar (IYK98, FPH00b)         \\
11& 511 &  $^4$28 & -- & 00 49 30.7 & -73 31 09 &  1.8 &   011 & 1.2\expo{35}:&   302 &   -- &  22.2 &   --  & --    &   --  & Be/X?                                           \\ 
12& 468 &  -- & -- & 00 49 43.8 & -73 23 02 & 14.9 &   001 & 2.7\expo{34}:&   315 &   -- &  14.7 & 17.7: & --    & 15.4: & Be/X?                                           \\ 
13& 444 &  34 & -- & 00 50 44.3 & -73 15 58 &  4.9 &   023 & 1.8\expo{36} &   387 &   -- &   6.9 & 15.41 & 15.44 &   --  & Be/X AXJ0051-733, 323.2 s pulsar (CSM97, YK98b, SC98, CO00)\\ 
14& 514 &  -- & -- & 00 50 46.8 & -73 32 47 & 33.7 &   010 & 2.4\expo{34}:&   393 &   -- &   9.2 &   --  & --    &   --  & Be/X?                                           \\ 
15& 421 &  36 & -- & 00 50 56.5 & -73 10 09 &  4.5 &   013 & 1.1\expo{35}:&   414 &   -- &   3.5 &   --  & --    &   --  & Be/X?                                           \\ 
16&  -- &  37 & -- & 00 50 56.9 & -72 13 31 &  1.4 &    -- & 2.9\expo{37} &   413 &   -- &   2.7 & 15.8: & --    & 16.4: & Be/X AXJ0051-722 91.12 s pulsar (CML98b, L98, SCB99)\\
17& 349 &  -- & -- & 00 51 19.5 & -72 50 43 & 15.6 &   001 & 3.6\expo{34}:&   447 &   -- &  14.6 &   --  & --    &   --  & Be/X?                                           \\ 
18&  99 &  -- & -- & 00 51 51.4 & -71 59 50 & 47.7 &   001 &     --       &   502 &    9 &  22.3 & 14.6: & --    & 13.3: & AC?                                                 \\ 
19& 424 &  41 & 25 & 00 51 51.5 & -73 10 31 &  2.2 &   055 & 4.7\expo{35}:&   504 &   -- &   4.2 & 13.1: & 14.4  & 14.4: & Be/X RXJ0051.9-7311 (CSM97, SCC99)\\ 
20& $^3$265 &--&27 & 00 51 53.1 & -72 31 50 &  2.0 &   112 & 1.4\expo{36} &   506 &   -- &   1.3 & 13.12 & 13.4  &   --  & Be/X RXJ0051.8-7231, 8.9 s pulsar (ISA97, SCB99)       \\ 
21& --  &  -- & 26 & 00 51 54.2 & -72 55 36 & 40.0 &    -- & 6.0\expo{34} &   521 &   -- &  28.6 &   --  & --    &   --  & Be/X?\\
22&  -- &  43 & -- & 00 52 05.4 & -72 25 55 & 15.8 &    -- & 5.6\expo{37} &   531 &   -- &   7.1 & 15.0  & --    & 15.4: & Be/X SMC X-3 \\
23& 453 &  44 & 29 & 00 52 13.9 & -73 19 13 &  1.9 &   011 & 1.3\expo{37} &   552 &   -- &   5.1 &   --  & 14.62 & 14.54 & Be/X RXJ0052.1-7319, 15.3 s pulsar (LPM99, ISC99)       \\ 
24& --  &  -- & 31 & 00 52 52.7 & -72 48 22 & 40.0 &    -- & 1.6\expo{35} &   618 &   -- &   7.8 & 14.28 & 14.28 & 15.6: & Be/X 2E0051.1-7304, AzV138 (GH85)\\
25&  94 &  $^4$46 & 32 & 00 52 54.7 & -71 58 08 &  2.0 &   004 & 2.0\expo{37} &   623 &   -- &  13.2 & 15.39 & 15.46 & 14.7: & Be/X RXJ0052.9-7158 (CSM97,SCC99) =? XTEJ0054-720 169.3 s pulsar\\ 
26& 246 &  $^4$48 & -- & 00 53 24.1 & -72 27 14 &  2.1 &   001 & 6.4\expo{34}:&   667 &   -- &  31.0 & 17.0: & --    & 15.8: & Be/X?                                           \\ 
27& 242 &  -- & 34 & 00 53 53.3 & -72 27 01 & 25.7 &   012 & 7.4\expo{36} &   717 &   -- &  22.7 &   --  & --    &   --  & Be/X? XTEJ0053-724, 46.63 s pulsar (CML98b)             \\ 
28& 547 &  -- & -- & 00 54 30.8 & -73 40 55 & 10.5 &   011 & 8.4\expo{37} &    -- &   -- &   --  & 15.7  & 16.0  &   --  & Be/X SMCX-2 (KP96)                              \\ 
29& 248 &  -- & -- & 00 54 33.1 & -72 28 08 & 45.7 &   001 & 1.5\expo{36}:&   772 &   -- &  26.6 &   --  & --    &   --  & Be/X? [hard]                                    \\ 
30& 324 &  57 & -- & 00 54 55.3 & -72 45 06 &  5.6 &   012 & 3.0\expo{35}:&   809 &   -- &   4.7 & 16.8: & --    & 14.8: & Be/X?                                           \\ 
31& 241 &  58 & 35 & 00 54 55.4 & -72 26 46 &  3.5 &   023 & 3.0\expo{37} &   810 &   -- &   3.1 & 15.24 & 15.28 &   --  & Be/X XTEJ0055-724, 59.07 s pulsar (ML98, SCI98, SCB99, CO00)\\ 
\noalign{\smallskip}
\hline
\end{tabular}
\\$^1$ B denotes 11 detections
\\$^2$ Entry in catalogue of MB99
\\$^3$ Parameters from observation 600453 when source was bright
\\$^4$ HRI position and error
\\$^5$ ASCA position and error
\label{tab-ma93}
\end{table}
\addtocounter{table}{-1}
\begin{table}
\caption[]{Continued}
\scriptsize
\begin{tabular}{rrrrccrrcrcrlllp{75mm}l}
\hline\noalign{\smallskip}
1& 2~  &3~    &4~ &     5      &      6    & 7~   &  8~   &   9          &  10   & 11   &12     &  ~13  &  ~14  &  ~15  & ~~~~16     \\
\hline\noalign{\smallskip} 
 &No  &  No & No &  RA        &  Dec      &\perr & det   & L$_{\rm x}$  & No    & T    & dist  &   ~B  &   ~V  &   ~R  & Remarks    \\
 &RP  &  RH & EI &&\hspace{-15mm}(J2000.0)&[\arcsec]&    & erg s$^{-1}$ & MA    &      & [\arcsec]&    &       &       &\\
\noalign{\smallskip}\hline\noalign{\smallskip}
32& --  &  -- & 38 & 00 56 03.3 & -72 21 32 & 40.0 &    -- & 4.0\expo{34} &   904 &   -- &  29.6 & 15.7: & --    & 14.3: & Be/X?\\
33& 270 &  -- & -- & 00 57 15.4 & -72 33 38 & 25.9 &   000 &     --       &   993 &   -- &  27.1 & 17.2: & --    & 16.7: & ?                                                    \\ 
34& 117? &  71? & 40 & 00 57 32.4 & -72 13 17 & 40.0 &    -- &     --       &  1021 &   -- &  23.6 & 15.4: & --    & 13.7: & ?\\
35& 114 &  73 & 41 & 00 57 48.4 & -72 02 42 &  7.9 &   001 & 1.6\expo{36} &  1036 &   -- &   9.5 & 16.8: & --    & 17.7: & Be/X? AXJ0058-720, 280.4 s pulsar (YK98b)               \\ 
36& 136 &  74 & -- & 00 57 50.1 & -72 07 56 &  5.1 &   012 & 4.3\expo{35}:&  1038 &   -- &   1.3 & 16.7: & --    & 15.8: & Be/X? [hard]                                    \\ 
37&  87 &  -- & -- & 00 57 59.5 & -71 56 37 & 19.2 &   001 & 5.7\expo{34}:&  1044 &   -- &  21.1 & 17.1: & --    & 15.5: & Be/X?                                           \\ 
38&  -- &  76 & -- & 00 58 12.9 & -72 30 45 &  3.1 &    -- & 2.1\expo{35}:&    -- &   -- &    -- & 16.7: & 14.9  & 15.5:  & Be/X RXJ0058.2-7231 (SCC99)\\
39&  47 &  79 & 43 & 00 58 37.2 & -71 35 50 &  1.3 &$^1$B30&     --       &  1083 &    5 &   1.8 & 16.8: & --    & 15.3: & SSS 1E0056.8-7154, PN LIN333 (KP96)                  \\ 
40&  53 &  -- & -- & 00 59 11.3 & -71 38 45 &  2.8 &   111 & 5.0\expo{37} &$^2$179&   -- &   10.1& 14.21 & 14.08 & 14.03 & Be/X RXJ0059.2-7138, 2.763 s soft pulsar (H94, SC96)     \\ 
41&  51 &  -- & -- & 00 59 41.7 & -71 38 15 & 14.5 &   300 &     --       &  1159 &    5 &   5.7 & 16.8: & --    & 16.5: & SSS? RXJ0059.6-7138, PN LIN357 (KPFH99)              \\ 
42& 132 &  93 & -- & 01 01 01.1 & -72 06 57 &  2.8 &   011 & 1.3\expo{36} &  1240 &   -- &   8.5 &   --  & --    &   --  & Be/X RXJ0101.0-7206 (KP96, SCB99)                \\ 
43& 159 &  $^4$95 & -- & 01 01 20.5 & -72 11 18 &  1.6 &   013 & 6.6\expo{35}:&  1257 &   -- &   4.7 &   --  & --    &   --  & Be/X? [nonstar]                                 \\ 
44& 121 &  96 & 46 & 01 01 37.0 & -72 04 19 &  7.2 &$^1$06B& 3.8\expo{35}:&  1277 &   -- &   5.5 & 17.7: & --    & 16.4: & Be/X?                                           \\ 
45& 220 &  97 & -- & 01 01 51.0 & -72 23 26 &  8.1 &   025 & 2.2\expo{35}:&  1288 &   -- &   9.8 & 14.7: & --    & 14.3: & Be/X?                                          \\ 
46&  92 &  -- & -- & 01 02 51.3 & -71 57 43 & 18.2 &   000 &     --       &  1338 &   -- &  14.1 & 15.1: & --    & 13.6: & ?                                                     \\ 
47& --  &  -- & 49 & 01 03 06.9 & -72 32 59 & 40.0 &    -- &     --       &  1357 &    5 &   4.8 &   --  & --    &   --  & ?\\
48&  77 &  -- & -- & 01 03 07.1 & -71 51 47 & 11.8 &   001 &     --       &  1365 &   -- &  14.6 & 18.8: & --    & 17.8: & ?                                                    \\ 
49& 143 & 101 & 50 & 01 03 14.0 & -72 09 16 &  3.4 &   067 & 1.5\expo{36} &  1367 &   -- &   1.1 & 14.73 & 14.80 & 14.74 & Be/X SAXJ0103.2-7209, 345.2 s pulsar (ISC98, HS94, CO00)  \\ 
50& 106 & 105 & -- & 01 03 37.0 & -72 01 39 &  5.0 &   013 & 3.0\expo{35}:&  1393 &   -- &   7.2 &   --  & --    &   --  & Be/X?                                           \\ 
51& 317 &  -- & -- & 01 04 07.4 & -72 43 59 & 17.7 &   002 & 3.8\expo{34}:&  1440 &   -- &   9.0 & 14.1: & --    & 14.4: & Be/X? or AGN? 13cm                              \\ 
52&  -- & 108 & -- & 01 04 35.6 & -72 21 43 &  2.3 &    -- & 4.8\expo{34}:&  1470 &   -- &   4.0 & 14.8: & --    & 15.1: & Be/X?\\
53& 163 & 110 & -- & 01 05 08.9 & -72 11 44 &  6.6 &   012 & 1.5\expo{35} &  1517 &   -- &   7.7 &   --  & --    &   --  & Be/X? AXJ0105-722, 3.343 s pulsar (YK98c, FHP00a)        \\ 
54& 737 &  -- & -- & 01 05 42.3 & -72 26 15 & 15.7 &   001 & 1.8\expo{34}:&  1544 &   -- &  12.8 & 14.2: & --    & 14.0: & Be/X?                                           \\ 
55& 120 &  -- & -- & 01 05 54.8 & -72 03 54 &  5.5 &   011 & 6.5\expo{34}:&  1557 &   -- &   3.7 &   --  & --    &   --  & Be/X?                                           \\ 
56& 279 &  -- & 56 & 01 07 10.9 & -72 35 36 & 11.1 &   001 & 2.3\expo{34}:&  1619 &   -- &  10.1 & 16.6: & --    & 15.4: & Be/X?                                           \\ 
57& 253 &  -- & -- & 01 09 01.2 & -72 29 07 & 17.6 &   012 &     --       &  1682 &    5 &  12.3 &   --  & --    &   --  & ? [hard]                                             \\ 
58& $^5$446 &  -- & -- & 01 11 14.5 & -73 16 50 & 30.0 &    -- & 2.0\expo{38} &    -- &   -- &    -- & 15.24 & 15.32 & 15.37 & Be/X XTEJ0111.2-7317, 31.03 s pulsar (CLC98a, WF98, ISC99, CHR99)\\
59& 506 &  -- & -- & 01 17 41.4 & -73 30 49 &  0.6 &   122 & 1.2\expo{38} &  1845 &   -- &   4.3 & 14.1: & 14.2  & 14.7: & Be/X RXJ0117.6-7330, 22.07 s pulsar (MFH99, CRW97)       \\ 
60& 501 &  -- & -- & 01 19 37.6 & -73 30 06 & 12.9 &   001 & 1.5\expo{34}:&  1867 &   -- &   8.3 & 15.1: & --    & 15.8: & Be/X?                                           \\ 
\noalign{\smallskip}
\hline
\noalign{\smallskip}
\end{tabular}

References:\\
(CHR99)  \cite{chr99}  1999,      (CLC98a) \cite{clc98a} 1998a,
(CML98b) \cite{cml98b} 1998b,     (CO00)   \cite{co00}   2000,
(CRW97)  \cite{crw97}  1997,      (CSM97)  \cite{csm97}  1997,
(FHW98)  \cite{fhw98}  1998,      (FHP00a) \cite{fhp00a} 2000a,
(FPH00b) \cite{fph00b} 2000b,     (GH85)   \cite{gh85}   1985,
(H94)    \cite{h94}    1994,      (HS94)   \cite{hs94}   1994,
(ISA97)  \cite{isa97}  1997,      (ISC98)  \cite{isc98}  1998,
(ISC99)  \cite{isc99}  1999,      (IYK98)  \cite{iyk98}  1998,
(KP96)   \cite{kp96}   1996,      (KP98)   \cite{kp98}   1998,
(KPFH99) \cite{kpfh99} 1999,      (L98)    \cite{l98}    1998,
(LPM99)  \cite{lpm99}  1999,      (M92)    \cite{m92}    1992,
(MFH99)  \cite{mfh99}  1999,      (ML98)   \cite{ml98}   1998,
(RLG94)  \cite{rlg94}  1994,      (SC96)   \cite{sc96}   1996,
(SC98)   \cite{sc98}   1998,      (SCB99)  \cite{scb99}  1999,
(SCC99)  \cite{scc99}  1999,      (SCI98)  \cite{sci98}  1998,
(WF98)   \cite{wf98}   1998,      (YK98a)  \cite{yk98a}  1998a,
(YK98b)  \cite{yk98b}  1998b,     (YK98c)  \cite{yk98c}  1998c
\end{table}

\end{document}